\documentclass[12pt]{article}
\usepackage{amsfonts}
\usepackage{amsmath}
\usepackage{amssymb}
\usepackage{graphicx}
\usepackage{color}
\usepackage[all, knot]{xy}
\usepackage{tikz}
\usepackage{array}
\usepackage{hyperref}
\usepackage{ulem}
\usepackage{booktabs}
\usepackage{needspace}

\usepackage{comment}

\usepackage[utf8]{inputenc}
\usepackage{epstopdf}
\usepackage{subcaption}
\usepackage{amsthm}
\usepackage{enumitem}
\usepackage{mathrsfs}
\usepackage{mathtools}

\usepackage[margin=3cm]{geometry}

\def \be {\begin{equation}}
\def \ee {\end{equation}}
\def \bea {\begin{eqnarray}}
\def \eea {\end{eqnarray}}

\def \rr {\raise.35ex\hbox{\small $\prime$}\kern-.17em{\mbox{\large $\imath$}}}

\def \dels {\partial\kern-.6em /\kern.1em}
\def \As {{A\kern-.5em / \kern.5em}}
\def \Ds {D\kern-.7em / \kern.5em}

\def \ks {k\kern-.5em /}
\def \ls {l\kern-.5em /}

\def \one {{\bf 1}}

\newcommand{\ci}[1]{}

\newcommand{\ba}{\begin{eqnarray}}
\newcommand{\ea}{\end{eqnarray}}
\newcommand{\bal}{\begin{align}}
\newcommand{\eal}{\end{align}}
\newcommand{\bay}[1]{\left(\begin{array}{#1}}
\newcommand{\eay}{\end{array}\right)}

\newcommand{\GM}{\mathrm{GM}}

\newcommand{\Prob}{\mathbb P}
\newcommand{\EE}{\mathbb E}
\newcommand{\btheta}{\boldsymbol\theta}
\newcommand{\OO}{\mathcal O}
\newcommand{\hide}[1]{}

\DeclareMathOperator{\Tr}{Tr}

\newlist{axioms}{enumerate}{2}
\setlist[axioms,1]{label=\textbf{A\arabic{axiomsi}.}, ref=A\arabic{axiomsi}}
\setlist[axioms,2]{label=\textbf{A\arabic{axiomsi}\rlap{\myEnumCounter{axiomsii}}.},%
                   ref=A\arabic{axiomsi}\myEnumCounter{axiomsii},%
                   align=parleft,%
                   leftmargin=0em,%
                   itemsep=1.4ex,%
                   before={\stepcounter{axiomsi}}}

  \usetikzlibrary{decorations.markings}

\begin{document}
\begin{titlepage}

\begin{center}

\textbf{\LARGE
Negative Junction Logarithms in\\ 
Scalar Multipartite Entanglement
\vskip.3cm
}
\vskip .5in
{\large
Chen-Te Ma$^{a}$ \footnote{e-mail address: yefgst@gmail.com}, 
\\
\vskip 1mm
}
{\sl
$^a$
Hetao Institute of  Mathematics and Interdisciplinary Sciences, Shenzhen,\\ 
Guangdong 518017, China.
}
\\
\vskip 1mm
\vspace{40pt}
\end{center}

\begin{abstract}
\noindent
A three-region junction in a free real relativistic scalar vacuum in $2+1$ dimensions yields a strictly negative logarithmic contribution to genuine tripartite R\'enyi multi-entropy for every integer $n\geq3$ and all nondegenerate sector angles. 
The replica subtraction exactly removes loop classes associated with a single interface. 
Surviving loops have positive, scale-independent weight, yielding a logarithm whose sign is fixed by the negative R\'enyi prefactor. 
The selector is empty at $n=2$.
The flat-sheet Friedrichs prescription fixes the scalar operator at the nonmanifold replica apex.
It yields a convergent angular spectral formula for the coefficient.
\end{abstract}
\end{titlepage}
\setcounter{tocdepth}{3}
{\hypersetup{linkcolor=black}\tableofcontents}
\newpage
\section{Introduction}
\label{sec:1}
\noindent
Replica methods glue copies (sheets) to compute density-matrix invariants~\cite{Holzhey:1994we}.
In two-dimensional conformal field theories (CFTs), twist operators implement this sewing~\cite{Lunin:2000yv} for disjoint intervals~\cite{Caraglio:2008pk,Calabrese:2009ez} and partial-transpose moments~\cite{Calabrese:2012nk}.
Bipartition corners yield angle-dependent logarithms in $2+1$ dimensions~\cite{Casini:2006hu,Helmes:2016fcp} with exact free-field results~\cite{Elvang:2015jpa} and twist-dimension relations~\cite{Bueno:2015qya}.
At a three-region junction, subtracting bipartition entropies from a multipartite replica observable removes interface contributions~\cite{Gadde:2022cqi,Penington:2022dhr}.
The question is what fixes the sign of the surviving junction logarithm for arbitrary sector angles and integer replica index.
Junction localization occurs in gapped free fermions~\cite{Iizuka:2026qqg} and gravitational duals of confining theories~\cite{Iizuka:2026ahd}.
Negative genuine multi-entropies are already known in several bosonic models~\cite{Harper:2025uui,Berthiere:2025toi}.
At R\'enyi index two, it vanishes for every pure bosonic Gaussian state~\cite{Camargo:2026xrd}.
For $n\geq3$, the replica cone's spherical cross-section, or angular link, has handles.
Its apex is therefore nonmanifold, so we need a scalar-operator prescription.
\\

\noindent
For the free real scalar in $2+1$ dimensions, we prove that the junction logarithm is strictly negative for every integer $n\geq3$ and all positive sector angles.
Replica subtraction cancels loops compatible with one interface, retaining positive junction-sensitive weight.
Scale invariance gives equal weight per logarithmic interval of loop size, producing a logarithm with negative R\'enyi prefactor.
This fixes the sign without solving the angular spectrum.
This extends loop representations for R\'enyi-2 information~\cite{Maghrebi:2015goa} and irreducible Casimir contributions~\cite{Schaden:2010wv,Gies:2003cv} to multipartite sewing, with local additivity and a convergent spectral formula.

\section{Observable and Prescription}
\label{sec:2}
\noindent
We consider the vacuum of a real scalar $\phi\in\mathbb R$, without periodic identification, with Euclidean action
\begin{equation}
I[\phi]=\frac12\int_{\mathbb R^D}d^D x\,
\bigl[(\partial\phi)^2+m^2\phi^2\bigr],\qquad m>0.
\label{eq:action}
\end{equation}
Here $D=d+1$ for positive integer spatial dimension $d$, $x^0$ is Euclidean time, $(\partial\phi)^2=\sum_{\mu=0}^{D-1}(\partial_\mu\phi)^2$, and $\Delta=\sum_{\mu=0}^{D-1}\partial_\mu^2$.
The mass $m$ and length cutoff $\epsilon>0$ regulate long and short distances. 
Regions $A,B,C$ partition the pure-state system; their time-zero interfaces, including junctions, form the branch locus. 
For our geometries, Brownian paths almost surely avoid this locus (Appendix~\ref{sec:appA}).
For integer $n\geq2$, we label the $n^2$ sheets by $G_n=\mathbb Z_n^2$ with addition modulo $n$~\cite{Gadde:2024taa}.
Crossing toward positive Euclidean time translates the label by $g_A=(1,0)$, $g_B=(0,1)$, or $g_C=(0,0)$ on the respective region.
We denote the full, ordinary cyclic, and unreplicated partition functions by $Z_n^{(3)}$, $Z_{n,X}^{(2)}$, and $Z_1$.
The normalized sewing contractions are invariant under independent unitary basis changes in each party~\cite{Ma:2026wpv,Gadde:2025csh}.
Our normalization and genuine subtraction are~\cite{Iizuka:2025ioc,Iizuka:2025caq}
\begin{align}
S_n^{(3)}&=\frac{\ln Z_n^{(3)}-n^2\ln Z_1}{n(1-n)},\nonumber\\
S_n(X)&=\frac{\ln Z_{n,X}^{(2)}-n\ln Z_1}{1-n},\nonumber\\
\GM_n^{(3)}&=S_n^{(3)}-\frac12\sum_{X=A,B,C}S_n(X).
\label{eq:normalization}
\end{align}
Here $S_n(X)=(1-n)^{-1}\ln\Tr\rho_X^n$, with $\rho_X$ the reduced density matrix.
If $C$ is absent, $n$ identical cyclic covers give $Z_n^{(3)}=(Z_{n,A}^{(2)})^n$ and $S_n^{(3)}=S_n(A)$, so purity makes the subtraction vanish.
The factor $1/2$ likewise compensates for each interface appearing in both adjacent entropies. 
This signed subtraction is a geometric multipartite signal; a negative value does not contradict positivity of the individual entropies. 
It need not remove all two-party correlations or be monotone under local operations and classical communication~\cite{Gadde:2026msg}. 
Our subtraction differs from the three-qubit residual tangle~\cite{Coffman:1999jd}.
\\

\noindent
The common Friedrichs prescription fixes allowed field behavior at the singular set and makes diffusion compatible across the full cover and its cyclic quotients. 
On the regular cover $M^\circ$, we admit limits of smooth compactly supported fields in the norm $(\|\phi\|_2^2+q[\phi])^{1/2}$, where $q[\phi]=\int_{M^\circ}|\nabla\phi|^2d\mathrm{vol}$ (Appendix~\ref{sec:appA}). 
This Friedrichs prescription fixes a nonnegative self-adjoint $-\Delta$ without contact interactions at the branch locus. 
Other singular domains~\cite{Kirsten:2007ur,Herzog:2016bhv} and regional-algebra choices~\cite{Casini:2013rba,Ma:2015xes} define different entropy prescriptions. 
Proper time $s$, with units of length squared, parametrizes diffusion and differs from $x^0$.
All traces use $s\geq\epsilon^2$ and the same outer Dirichlet boundary, symmetric under $x^0\mapsto-x^0$, removed after combination.

\section{Exact Loop Selector}
\label{sec:3}
\begin{figure}[!t]
\centering
\includegraphics[width=\columnwidth]{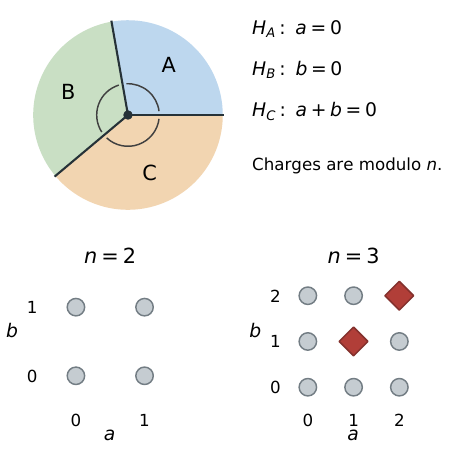}
\caption{Three spatial sectors and the exact monodromy selector.
Residues with $a=0$, $b=0$, or $a+b=0\pmod n$ cancel in Eq.~\eqref{eq:selector}.
Diamonds mark surviving charges: none at $n=2$; $(1,1)$ and $(2,2)$ at $n=3$.}
\label{fig:selector}
\end{figure}

\noindent
The Gaussian determinant organizes scalar fluctuations as diffusion loops.
The same loop in the original Euclidean space can close on one replica cover but end on another sheet on a different cover, allowing subtraction before integration.
The free kernel $p_s(x,y)=(4\pi s)^{-D/2}\exp(-|x-y|^2/(4s))$ describes diffusion with generator $\Delta$ and coordinate variance $2s$.
Conditioning on return gives a Brownian bridge $\gamma(0)=\gamma(s)=x$, an exact heat-kernel representation without additional assumptions about the vacuum~\cite{Baer:2011}.
The lifted path ends on the initial sheet translated by $h_n(\gamma)=(a,b)\in G_n$.
This replica-label displacement, called its charge or monodromy, remains well defined even when cut crossings are not isolated.
The ordinary cyclic covers retain the residues
\begin{equation}
\ell_A(a,b)=a,\qquad \ell_B(a,b)=b,\qquad
\ell_C(a,b)=-a-b.
\label{eq:quotients}
\end{equation}
Indeed, signed crossings of a smooth closed representative obey $N_A+N_B+N_C=0$, with $a=N_A$, $b=N_B$, and $\ell_C=N_C$ modulo $n$.
Identifying sheets differing by $H_X=\ker\ell_X$ leaves the usual $n$-sheet cyclic replica (Appendix~\ref{sec:appA}).
We define
\begin{equation}
\mathcal Q_n=G_n\setminus(H_A\cup H_B\cup H_C).
\label{eq:Q}
\end{equation}
Survival requires all three cyclic residues to be nonzero: none of the cyclic lifts returns to its initial sheet.
We write $k_s=p_s(x,x)$ and use $\Prob^s_{x,x},\EE^s_{x,x}$ for bridge probability and expectation.
In a common bounded Dirichlet domain $\Omega$, let $J_\Omega$ indicate survival and $k_h(s;x)$ be the massless Friedrichs kernel from a lift of $x$ to its sheet translate by $h$.
The kernel is lift-independent and nonnegative.
On the quotient $X$, endpoints differing by $h\in H_X$ are identified, so its return kernel sums over them without averaging (Appendix~\ref{sec:appB}).
The full cover requires $h=0$, whereas forgetting sheet labels allows every $h\in G_n$.
Thus, the traces $K(s)=\Tr \exp(s\Delta)$ obey
\begin{align}
\frac{K_n^{(3)}(s)}{n^2}&=\int_\Omega d^Dx\,k_0(s;x),\nonumber\\
\frac{K_{n,X}^{(2)}(s)}n&=\int_\Omega d^Dx\sum_{h\in H_X}k_h(s;x),\nonumber\\
K_1(s)&=\int_\Omega d^Dx\sum_{h\in G_n}k_h(s;x).
\label{eq:heat}
\end{align}
The factors $n^2,n,1$ count initial sheets; subtraction acts on shared nonnegative kernels, charge by charge.
Equivalently, we obtain $k_h=k_s\EE^s_{x,x}[J_\Omega\one_{\{h_n=h\}}]$, where $\one_E$ is the event indicator.
Real Gaussian integration gives $Z=[\det(-\Delta+m^2)]^{-1/2}$ and $\ln Z=\tfrac12\int_{\epsilon^2}^{\infty}ds\,\exp(-m^2s)K(s)/s$ with common normalization~\cite{Callan:1994py}.
We define
\begin{equation}
\mathcal K_n(s)=\frac{K_n^{(3)}(s)}{n^2}
-\frac1{2n}\sum_{X=A,B,C}K_{n,X}^{(2)}(s)+\frac12K_1(s).
\label{eq:combined-trace}
\end{equation}
Factoring out $1/(1-n)$ leaves $\ln Z_1$ coefficient $-n+3n/2=n/2$; including the determinant factor $1/2$ gives
\begin{equation}
\GM_n^{(3)}=-\frac{n}{2(n-1)}\int_{\epsilon^2}^{\infty}
\frac{ds}{s}\,e^{-m^2s}\mathcal K_n(s).
\label{eq:combined-entropy}
\end{equation}
For each loop, twice its weight in $\mathcal K_n$ is
\begin{equation}
2\one_{\{0\}}(h_n)-\sum_{X=A,B,C}\one_{H_X}(h_n)+1
=\one_{\mathcal Q_n}(h_n).
\label{eq:selector}
\end{equation}
Any two $H_X$ intersect only at zero, including for composite $n$.
Zero charge cancels with weight $2-3+1=0$; a nonzero charge closing on one cyclic cover cancels with $0-1+1=0$.
Only charges closing on none of the cyclic covers retain unit weight.
Applying the identity to each nonnegative $k_h$ gives $\mathcal K_n=W_n/2$ and
\begin{align}
\GM_n^{(3)}&=-\frac{n}{4(n-1)}
\int_{\epsilon^2}^{\infty}\frac{ds}{s}\,e^{-m^2s}W_n(s),
\label{eq:master}\\
W_n(s)&=\int_\Omega d^Dx\sum_{h\in\mathcal Q_n}k_h(s;x)\nonumber\\
&=\int_\Omega d^Dx\,k_s\,
\EE^s_{x,x}[J_\Omega\one_{\mathcal Q_n}(h_n)]\geq0.
\label{eq:W}
\end{align}
Eq.~\eqref{eq:master} makes the sign explicit: every retained weight is nonnegative, and the overall R\'enyi prefactor is negative.
The filter tests net sheet charge; visiting all three regions is insufficient because opposite crossings can cancel.
As $\Omega\uparrow\mathbb R^D$, monotone convergence removes the boundary, with $J_\Omega\to1$ and $k_h$ the increasing-domain limit.
This sign identity holds for other tripartition geometries and dimensions when compatible operator prescriptions and almost-sure branch avoidance hold.
Each subgroup has $n$ elements, so inclusion--exclusion gives
\begin{equation}
|\mathcal Q_n|=n^2-(3n-3+1)=(n-1)(n-2).
\label{eq:count}
\end{equation}
At $n=2$, the empty selector gives $\GM_2^{(3)}=0$, consistent with the Gaussian theorem without excluding multipartite correlations.
At $n=3$, loop reversal exchanges the survivors $(1,1)$ and $(2,2)$ (Fig.~\ref{fig:selector}).

\section{Strict Junction Logarithm}
\label{sec:4}
\noindent
A strictly negative logarithmic coefficient requires three further facts: the surviving weight is finite, is nonzero, and is independent of diffusion scale. 
Take $D=3$ and positive sector angles $\btheta=(\theta_A,\theta_B,\theta_C)$ summing to $2\pi$.
The boundaries form $Y=\bigcup_{i=1}^3\{ru_i:r\geq0\}$, where the $u_i$ are unit vectors in the time-zero plane.
A small oriented loop encircling one ray, called a meridian, crosses adjacent sectors oppositely and carries the sewing-translation difference. 
The three charges $(1,0),(-1,1),(0,-1)$ generate $H_B,H_C,H_A$, so a selected loop cannot stay near a single ray.
\\

\noindent
For the Euclidean ball $B(x,r)=\{y:|y-x|<r\}$, Appendix~\ref{sec:appC} constructs $\eta>0$ so $B(x,\eta|x|)$, $x\ne0$, avoids the apex and meets at most one ray. 
A selected loop must leave this ball to detect more than one ray, making some coordinate excursion at least $q=\eta|x|/\sqrt3$.
Reflection bounds each coordinate's two-sided bridge excursion by $2\exp(-q^2/s)$; summing over three coordinates gives
\begin{equation}
\Prob^s_{x,x}[h_n\in\mathcal Q_n]
\leq \min\{1,6e^{-\eta^2|x|^2/(3s)}\}.
\label{eq:tail}
\end{equation}
The Gaussian suppression controls loops based far from the junction; integration gives the finite bound $W_n(s)\leq9\sqrt3/(4\eta^3)$ uniformly in $s,n$. 
The mass enters only through $\exp(-m^2s)$ in Eq.~\eqref{eq:master}; $W_n$ is the massless geometric weight. 
Rescaling $x=\sqrt{s}\,y$ cancels the return kernel's $s^{-3/2}$, while dilation preserves the rays and charge, giving 
\begin{align}
W_n(s)&=\mathcal A_n(\btheta),\nonumber\\
\mathcal A_n&=(4\pi)^{-3/2}\int_{\mathbb R^3}d^3y\,
\Prob^1_{y,y}[h_n\in\mathcal Q_n].
\label{eq:A}
\end{align}
The meridians generate $G_n$, so the cover is connected and diffusion reaches every regular sheet with positive kernel $k_h(s;x)$ for $s>0$ (Appendix~\ref{sec:appD}). 
For $n\geq3$, $(1,1)\in\mathcal Q_n$, so integrating this positive kernel over any regular base-point ball proves $\mathcal A_n>0$. 
The dimensionless $\mathcal A_n$ is an integrated return weight, not a probability. 
Each logarithmic interval of diffusion length $L_s=\sqrt{s}$ contributes equally because $ds/s=2\,dL_s/L_s$. 
The range $\epsilon\ll L_s\ll m^{-1}$ produces the logarithm; the mass suppresses larger loops.
\\

\noindent
Setting $u=m^2s$ in Eq.~\eqref{eq:master} gives the exact sharp-cutoff expression
\begin{equation}
\GM_n^{(3)}=-\frac{n\mathcal A_n(\btheta)}{4(n-1)}
E_1(m^2\epsilon^2),\qquad
E_1(z)=\int_z^\infty\frac{e^{-u}}u\,du.
\label{eq:E1}
\end{equation}
For $z>0$, we get $E_1(z)=-\gamma_{\rm E}-\ln z+\OO(z)$ as $z\downarrow0$, where $\gamma_{\rm E}$ is Euler's constant (Appendix~\ref{sec:appD}).
Using $-\ln(m^2\epsilon^2)=2\ln[1/(m\epsilon)]$ yields
\begin{align}
\GM_n^{(3)}&=b_n(\btheta)\ln\frac1{m\epsilon}+\OO(1),\nonumber\\
b_n(\btheta)&=-\frac{n}{2(n-1)}\mathcal A_n(\btheta),
\qquad b_n<0\quad(n\geq3).
\label{eq:main}
\end{align}
Here $\OO(1)$ remains bounded as $\epsilon\to0$ at fixed $m$ and angles.
Thus, $b_n$ depends on the angles and replica index, while $m$ and $\epsilon$ set the infrared and ultraviolet scales. 
At fixed operator, $ds/s$-integrable changes of a common scaling-cutoff profile preserve $b_n$ and alter only the bounded term (Appendix~\ref{sec:app-cutoff}). 
Related two-dimensional CFT analyses separate universal logarithms from boundary-sensitive contributions~\cite{Ohmori:2014eia,Ma:2016pah,Ma:2016deg}.
The full $E_1$ expression remains cutoff dependent.

\section{Finite Partitions}
\label{sec:5}
\noindent
The logarithmic divergence is local: sufficiently small loops resolve one junction at a time. 
We consider a compact straight-edge interface graph with two-party corners and a finite set $\mathcal V_3$ of triple junctions having positive angles $\btheta_v$. 
Near each junction, the partition matches a three-sector cone; distinguishing them requires a fixed-distance excursion, exponentially unlikely at short time.
Elsewhere, locally confined loops encounter only two labels, so $N_P+N_Q=0$ gives $h_n=N_P(g_P-g_Q)$ in a canceled subgroup, including at corners.
Integrating excursion bounds over junction neighborhoods, the bounded complement, and the exterior (Appendix~\ref{sec:appG}) yields
\begin{align}
W_n(s)&=\sum_{v\in\mathcal V_3}\mathcal A_n(\btheta_v)
+\OO(e^{-c\ell_0^2/s}),\qquad s\downarrow0,\nonumber\\
\GM_n^{(3)}&=\left[\sum_{v\in\mathcal V_3} b_n(\btheta_v)\right]\ln\frac{\ell}{\epsilon}+\OO(1).
\label{eq:additive}
\end{align}
Here $c,\ell_0>0$ depend on the fixed geometry, $\ell>0$ is any reference length, and $m>0$ is fixed. 
The exponentially small remainder and mass correction $\exp(-m^2s)-1=\OO(m^2s)$ give finite proper-time integrals near $s=0$. 
Only the sum of local constant weights multiplies the logarithm; changing $\ell$ shifts the bounded term. 
For $n\geq3$, every triple junction contributes negatively. 
Without junctions, surviving loops must span separated interfaces and can give only a finite, nonlocal signal.

\section{Replica Link and Spectral Coefficient}
\label{sec:6}
\noindent
The coefficient's sign is fixed; the angular spectrum determines its magnitude.
The unit-radius link $\Sigma_n^{(3)}$ is an $n^2$-sheeted cover of the round sphere, branched at the three ray directions $u_i$ fixed by the sector angles.
The meridians generate $G_n$, making it connected.
Each meridian has $n$ cycles of length $n$, giving $n$ branch points of degree $n$ above each branch value.
Each point contributes $n-1$ to the total branching correction $3n(n-1)$; the Riemann--Hurwitz formula gives the genus (handle count) $g_n$:
\begin{equation}
2-2g_n=2n^2-3n(n-1),\qquad
2g_n=(n-1)(n-2)=|\mathcal Q_n|.
\label{eq:genus}
\end{equation}
The $n=3$ link is a torus~\cite{Harper:2024ker}; related covers and singular links appear in Refs.~\cite{Ma:2026frw,DelZotto:2026fpw}.
An ordinary point has a spherical link; persistent handles make this apex nonmanifold for $n\geq3$ (Appendix~\ref{sec:appF}).
Diffusion avoids the apex but detects the rays through winding.
The Friedrichs prescription excludes defect interactions; the selector fixes the sign while angular propagation fixes the magnitude.
\\

\noindent
Let $K_\Sigma(t)=\Tr\exp(t\Delta_\Sigma)$, $t>0$, be the link's Friedrichs heat trace; $\Sigma_{n,X}^{(2)}$ denotes its cyclic quotients.
If $0=\lambda_{\Sigma,0}<\lambda_{\Sigma,1}\leq\cdots$ are the eigenvalues of $-\Delta_\Sigma$ counted with multiplicity, then $K_\Sigma(t)=\sum_{j=0}^{\infty}\exp(-t\lambda_{\Sigma,j})$.
Each quotient branches at the two rays bounding $X$; the third meridian lies in $H_X$ and acts trivially.
The links inherit the spherical metric away from branching and are completed at the branch points.
All use the same Friedrichs prescription (Appendix~\ref{sec:appH}).
The weights of Eq.~\eqref{eq:combined-trace} give the link combination:
\begin{equation}
F_n(t)=\frac{K_{\Sigma_n^{(3)}}(t)}{n^2}
-\frac1{2n}\sum_{X=A,B,C}K_{\Sigma_{n,X}^{(2)}}(t)
+\frac12K_{S^2}(t).
\label{eq:F}
\end{equation}
The same selector makes $F_n$ half the total return weight of surviving angular loops.
With spherical kernel $k_t^{S^2}$ and bridge measure $\Prob^{t,S^2}_{\omega,\omega}$, it gives
\begin{equation}
F_n(t)=\frac12\int_{S^2}d\omega\,k_t^{S^2}(\omega,\omega)
\Prob^{t,S^2}_{\omega,\omega}[h_n\in\mathcal Q_n]\geq0.
\label{eq:spherical-selector}
\end{equation}
Here $d\omega$ is spherical area measure, $\int_{S^2}d\omega=4\pi$, and $t$ is dimensionless.
For an angular eigenvalue $\lambda\geq0$, writing $f(r)=v(r)/r$ reduces the radial operator to $-\partial_r^2+\lambda/r^2$, with Bessel index $\nu=\sqrt{\lambda+1/4}$~\cite{Bordag:1996fw}.
For each radial eigenmode, the Friedrichs condition selects the regular behavior $f(r)\propto r^{\nu-1/2}$ at the apex.
For the original mode $f$, the diagonal radial heat kernel relative to $r^2dr$ is
\begin{equation}
k_\nu(r,r;s)=\frac{e^{-z}I_\nu(z)}{2sr},
\qquad z=\frac{r^2}{2s}.
\label{eq:radial-kernel}
\end{equation}
Here $r,s,z>0$, $\nu\geq1/2$, and $I_\nu$ is the modified Bessel function defined in Eq.~\eqref{eq:app-bessel-definition}; Eq.~\eqref{eq:app-cone-radial-kernel} derives the kernel.
All cone traces use common radial damping $\exp\big(-(\cosh\alpha-1)z\big)$, $\alpha>0$, to regulate the infinite volume.
Using $r\,dr=s\,dz$ with the trace measure $r^2dr$ gives
\begin{equation}
\frac12\int_0^\infty dz\,e^{-z\cosh\alpha}I_\nu(z)
=\frac{e^{-\alpha\nu}}{2\sinh\alpha}.
\label{eq:radial-trace}
\end{equation}
We define $B_n(\alpha)$ by replacing each $K_\Sigma(t)$ in Eq.~\eqref{eq:F} with $\sum_{j=0}^{\infty}\exp[-\alpha\sqrt{\lambda_{\Sigma,j}+1/4}]$.
The nonnegative damped combination $B_n(\alpha)/(2\sinh\alpha)$ increases to $\mathcal A_n/2$ as $\alpha\downarrow0$.
Subordination, a positive superposition of ordinary heat traces, gives
\begin{align}
\mathcal A_n&=\lim_{\alpha\downarrow0}\frac{B_n(\alpha)}{\sinh\alpha},\nonumber\\
B_n(\alpha)&=\frac{\alpha}{2\sqrt\pi}
\int_0^\infty\frac{dt}{t^{3/2}}\,
e^{-\alpha^2/(4t)-t/4}F_n(t).
\label{eq:spectral-bridge}
\end{align}
Positivity permits monotone convergence under the integral; the prefactor $\alpha/\sinh\alpha$ tends to one, yielding
\begin{equation}
\boxed{\displaystyle
\mathcal A_n(\btheta)=\frac1{2\sqrt\pi}
\int_0^\infty\frac{dt}{t^{3/2}}\,e^{-t/4}F_n(t).}
\label{eq:spectral}
\end{equation}
The shift $1/4$ comes from the three-dimensional radial Laplacian, not an added curvature interaction.
Selected loops must span a fixed angular distance, making $F_n(t)$ exponentially small at short time.
At long time, each connected link has one zero mode, giving $F_n(t)\to(n-1)(n-2)/(2n^2)$; the factor $\exp(-t/4)$ makes this endpoint integrable.
Eq.~\eqref{eq:spectral} determines the magnitude once the angular spectra are supplied.

\section{Replica-Index Constraints}
\label{sec:7}
\noindent
Radial projection gives a thrice-punctured sphere times $(0,\infty)$, whose fundamental group is freely generated by two meridians.
Orient them with charges $(1,0),(0,1)$, the second reversing the earlier $(0,-1)$ meridian.
Summing signed windings defines $(a,b)\in\mathbb Z^2$ (abelianization); reducing modulo $n$ gives $h_n$.
With $\mid,\nmid$ denoting divisibility and its negation, survival means
\begin{equation}
n\nmid a,\qquad n\nmid b,\qquad n\nmid(a+b).
\label{eq:integer}
\end{equation}
For $N>n\geq2$ with $n\mid N$, multiples of $N$ are multiples of $n$, so survival at $n$ implies survival at $N$.
The residue $(n,1)$ at index $N\geq2n>n+1$ survives but reduces to a canceled residue modulo $n$.
Its strictly positive kernel weight (Appendix~\ref{sec:appD}) gives
\begin{equation}
-\frac{N-1}{N}b_N>
-\frac{n-1}{n}b_n,
\qquad n\mid N,\quad N>n\geq2.
\label{eq:divisibility}
\end{equation}
This ordering applies only when one index divides the other. 
For $n>\max\{|a|,|b|,|a+b|\}$, divisibility by $n$ reduces to equality to zero. 
Survival therefore converges pointwise to $a,b,a+b\neq0$ as integers; call its weight $\mathcal A_\infty$.
The common bound~\eqref{eq:tail} permits dominated convergence, giving
\begin{equation}
0\leq\mathcal A_n<\mathcal A_\infty<\infty,
\qquad
\lim_{n\to\infty}b_n=-\frac12\mathcal A_\infty<0.
\label{eq:infty}
\end{equation}
Strictness follows from $\mathcal A_\infty\geq\mathcal A_{2n}>\mathcal A_n$. 
Branch clearance (Appendix~\ref{sec:appA}) stabilizes almost every bridge's charge under small angle changes. 
Dominated convergence, using Eq.~\eqref{eq:tail} uniformly on compact angle sets, gives continuity for $\theta_X>0$, $\sum_{X=A,B,C}\theta_X=2\pi$. 
We assume no regularity at a vanishing sector angle.

\section{Outlook}
\label{sec:8}
\noindent
At a three-sector junction, the filter retains finite positive weight at every scale, fixing the negative logarithm; short-time localization makes its coefficient additive over separated junctions. 
The Friedrichs prescription makes the singular geometry calculable, while the compact-link spectra determine the coefficient's magnitude. 
These integer-replica results do not establish analytic continuation to $n=1$.

\section*{Acknowledgments}
\noindent 
The author thanks Ma-Ke Yuan for her helpful discussion and thanks Nan-Peng Ma for his encouragement.

\appendix
\section{Operator Domain and Replica Quotients}
\label{sec:appA}
\noindent
We use the common regulators of Sec.~\ref{sec:2} with $\Omega_R=\{|x|<R\}$ and regular cover $M_R^\circ=\pi^{-1}(\Omega_R\setminus\mathcal B)$.
Here $\pi$ forgets the sheet and $\mathcal B$ is the closed branch locus.
Smooth Dirichlet-domain exhaustion realizes the specified $C_c^\infty(M_R^\circ)$ gradient-form closure and its stopped-diffusion kernel.
\\

\noindent
The planar hitting probability is $f(r)=\ln(L/r)/\ln(L/\delta)$ for $0<\delta<r<L$.
Indeed, $f''+f'/r=0$ and $f(\delta)=1,f(L)=0$ describe hitting $\delta$ before $L$.
Taking $\delta\downarrow0$ and $L\uparrow\infty$ excludes hitting the origin.
Transverse projection therefore proves polarity of finite unions of three-dimensional supporting lines, hence of our rays and segments.
Starting on a supporting line but outside $\mathcal B$ causes no exception: at positive times positions have densities, and the Markov property applies along times decreasing to zero.
For an event $E$ determined before $u<s$,
\begin{equation}
\mathbb P^s_{x,x}(E)
=\frac{\mathbb E_x[\mathbf1_Ep_{s-u}(B_u,x)]}{p_s(x,x)}.
\label{eq:app-A-bridge-null-events}
\end{equation}
Here $B$ is a Brownian motion starting at $x$.
Thus, bridge conditioning preserves null events.
Taking $u\uparrow s$ and $x\notin\mathcal B$ proves almost-sure avoidance throughout.
Compactness then gives positive branch clearance and unique continuous path lifting.
General partitions require polarity as a hypothesis.
Adding one to the $C$ quotient's upper sheet labels changes $(-1,-1,0)$ into $(0,0,1)$, establishing Eq.~\eqref{eq:quotients}.

\section{Kernel Identity and Infinite-Volume Limit}
\label{sec:appB}
\noindent
We write $M_R$ for the full cover of $\Omega_R$ and $T_h$ for translation of sheets by $h\in G_n$.
For a subgroup $H\subset G_n$, we set $N_H=M_R/H$ and pull back by $P_Hf=f\circ\rho_H$, with $\rho_H(p)=[p]$.
All $L_M=-\Delta_M$ use the common Friedrichs prescription.
Both the squared field norm and the gradient energy acquire the factor $|H|$.
Finite-group averaging identifies the closed quotient form with the $H$-invariant cover form under $|H|^{-1/2}P_H$.
Consequently, we get $\exp(-sL_{M_R})P_H=P_H\exp(-sL_{N_H})$.
Unfolding this evolution into its $|H|$ preimages gives
\begin{equation}
k_{N_H}(s;[p],[q])=\sum_{h\in H}k_{M_R}(s;p,T_hq).
\label{eq:app-kernel-quotient}
\end{equation}
There is no averaging factor relative to the quotient volume.
Appendix~\ref{sec:appA} removes trivial-monodromy strata.
Deck invariance and abelianness of $G_n$ make the diagonal independent of the initial sheet; counting $n^2/|H|$ such sheets proves Eq.~\eqref{eq:heat}.
The maximum principle and Dirichlet exhaustion give pointwise kernel nonnegativity, not merely nonnegative spectrum.
\\

\noindent
Endpoint conditioning, tested against arbitrary endpoint functions, gives
\begin{equation}
k_{M_R}(s;(x,q),(y,q'))
=p_s(x,y)\mathbb E^s_{x,y}[J_R\mathbf1_{\{T_\gamma(q)=q'\}}],
\label{eq:app-B-cover-kernel}
\end{equation}
Here $q,q'\in G_n$ label sheets, $J_R$ indicates survival inside $\Omega_R$, and $T_\gamma(q)$ is the final lifted sheet.
We normalize the conditional measure by $p_s(x,y)$, as in Appendix~\ref{sec:appA}.
For the diagonal limit, couple bridges as $x+v(y-x)/s+W_v-vW_s/s$, $0\leq v\leq s$, where $W$ is Brownian motion started at zero.
Branch clearance stabilizes the sheet event; immediate normal Brownian fluctuations exclude boundary touching without crossing.
Dominated convergence therefore permits $y\to x$.
\\

\noindent
The selector yields $\mathcal K_{n,R}=W_{n,R}/2$, hence Eq.~\eqref{eq:master} with its real-scalar prefactor.
Extend the nonnegative spatial integrand by zero outside $\Omega_R$.
Bounded bridges give $J_R\uparrow1$ with unchanged charge.
Monotone convergence in expectation, position, and proper time proves the infinite-volume formula without subtracting divergent determinants.
Appendix~\ref{sec:appC} establishes three-ray finiteness; a divergent proper-time integral instead gives entropy limit $-\infty$.

\section{Three Rays: Finite Junction Weight}
\label{sec:appC}
\noindent
For three distinct planar unit directions $u_i$, we set $Y=\bigcup_{i=1}^3\{ru_i:r\geq0\}$ and
\begin{equation}
\delta=\min_{1\leq i<j\leq3}\arccos(u_i\cdot u_j)>0,
\qquad \eta=\min\{1/4,\sin(\delta/2)/4\}.
\label{eq:app-eta}
\end{equation}
For $x\neq0$, the ball $B(x,\eta|x|)$ excludes the apex.
If it met two rays, the radial coordinates would exceed $(1-\eta)|x|$, giving separation at least $2(1-\eta)|x|\sin(\delta/2)\geq6\eta|x|$, whereas the ball diameter is $2\eta|x|$.
A single intersecting ray is a straight chord; its complement has fundamental group $\mathbb Z$ generated by the canceled meridian.
With no ray, the ball is simply connected.
Thus, selected loops must leave the ball.
\\

\noindent
For a duration-$s$ one-dimensional bridge $\beta$ from zero to zero with generator $\Delta$, reflection at $q>0$ gives $\mathbb P(\max_{[0,s]}\beta\geq q)=p_s(2q)/p_s(0)=\exp(-q^2/s)$, using $p_s(z)=(4\pi s)^{-1/2}\exp[-z^2/(4s)]$.
Both signs and three coordinate thresholds $q/\sqrt3$ give excursion probability at most $6\exp[-q^2/(3s)]$.
Setting $q=\eta|x|$ proves Eq.~\eqref{eq:tail}; Gaussian integration gives
\begin{equation}
W_n(s)\leq6(4\pi s)^{-3/2}(3\pi s/\eta^2)^{3/2}
=9\sqrt3/(4\eta^3).
\label{eq:app-uniform-weight-bound}
\end{equation}
The bound is uniform in $n,s$ and on compact nondegenerate angle sets.
After removing the outer boundary, $s^{-1/2}\gamma(st)$, $0\leq t\leq1$, is a unit bridge with unchanged charge.
The substitution $x=\sqrt s\,y$ cancels the return-density factor $s^{-3/2}$, proving Eq.~\eqref{eq:A} for finite weights.

\section{Strict Positivity and Logarithmic Coefficient}
\label{sec:appD}
\noindent
We prove positivity, extract the logarithm, and establish invariance under common cutoff changes.
On the connected regular cover, enclose a path between points $p,q$ in a connected smooth domain $U$ with compact closure.
The strong maximum principle and domain comparison give $k_M(s;p,q)\geq k_U(s;p,q)>0$ for $s>0$, and Dirichlet exhaustion yields the full Friedrichs kernel (Appendix~\ref{sec:appA}).
For $n\geq3$, $(1,1)\in\mathcal Q_n$; continuity gives a positive minimum of $k_M(1;\widetilde y,T_{(1,1)}\widetilde y)$ over a lifted closed base ball avoiding the rays.
Integrating this selected term proves $\mathcal A_n>0$ while $\mathcal Q_2=\varnothing$ gives $\mathcal A_2=0$.
\\

\noindent
Appendix~\ref{sec:appC} bounds the integral uniformly in outer volume at fixed $m,\epsilon>0$.
After removing the boundary, $W_n(s)=\mathcal A_n$; substituting $u=m^2s$ gives Eq.~\eqref{eq:E1}.
Separating the logarithmic term gives
\begin{equation}
E_1(z)=-\gamma_{\rm E}-\ln z+R(z),
\qquad R(z)=\int_0^z\frac{1-e^{-u}}u\,du.
\label{eq:app-E1-remainder}
\end{equation}
Here $z>0$, $\gamma_{\rm E}$ is Euler's constant, and $0\leq R(z)\leq z$ due to $0\leq1-\exp(-u)\leq u$.
Then $-\ln(m^2\epsilon^2)=2\ln[1/(m\epsilon)]$ gives Eq.~\eqref{eq:main}.

\subsection{Common Cutoff Changes}
\label{sec:app-cutoff}
\noindent
For bounded measurable $\chi:(0,\infty)\to\mathbb R$, require $\int_0^\infty|\delta\chi(u)|\,du/u<\infty$, where $\delta\chi=\chi-\mathbf1_{[1,\infty)}$.
We apply the cutoff to the combined trace; separate determinants require a common auxiliary cutoff until combination.
As $m\epsilon\to0$, dominated convergence gives $\int_0^\infty\exp(-m^2\epsilon^2u)\delta\chi(u)\,du/u\to c_\chi:=\int_0^\infty\delta\chi(u)\,du/u$.
The observable therefore shifts by $-n\mathcal A_n c_\chi/[4(n-1)]+o(1)$, preserving the logarithm at fixed operator, domain, and angles.

\section{Local Additivity for Finite Polygonal Partitions}
\label{sec:appG}
\noindent
Let the full-plane tripartition have a finite, compact, embedded straight-edge interface graph $\Gamma$, with positive edge lengths, intersections only at endpoints, degree at most three, and distinct labels on every edge.
Triple vertices $v_j$, $j=1,2,\ldots,J$, have three distinct labels and positive angles; other vertices have two-label neighborhoods.
Connectivity is not required; empty sums vanish.
We write $W_n^\Gamma$ for its infinite-volume weight in Eq.~\eqref{eq:W}.
\\

\noindent
The two-label cancellation in the Letter extends to bridges by smooth approximation away from the graph.
We choose $\ell>0$ with disjoint three-dimensional balls $B(v_j,4\ell)$ matching the local cones, and $R$ with $\Gamma\subset B(0,R/2)$ and $B_j:=B(v_j,\ell)\subset B(0,R)$.
A finite cover by half-sized two-label balls gives the compact complement $K=\overline{B(0,R)}\setminus\bigcup_jB_j$ a uniform two-label radius $\rho>0$.
The graph/cone selectors agree on $B_j$ unless the bridge travels $3\ell$; selected bridges on $K$ travel $\rho$, and those with $|x|>R$ travel $|x|/2$.
For $\tau=\ell^2/s\geq1$, Appendix~\ref{sec:appC} bounds the integrated junction, cone-tail, compact, and exterior errors:
\begin{align}
\left|W_n^\Gamma(s)-\sum_{j=1}^{J}\mathcal A_n(\boldsymbol\theta_j)\right|
&\leq C\left[\tau^{3/2}e^{-3\tau}+\sum_{j=1}^{J} e^{-a_j\tau}\right]\nonumber\\
&\quad+C\left[\tau^{3/2}e^{-\rho^2/(3s)}+e^{-R^2/(24s)}\right]\nonumber\\
&\leq C'e^{-c\tau}.
\label{eq:app-local-additivity}
\end{align}
Here $a_j=\eta_j^2/6$, with $\eta_j$ from Appendix~\ref{sec:appC} for cone $j$, and $C,c>0$ depend only on geometry, not $n$.
Splitting Gaussian exponents bounds the tails; absorbing $\tau^{3/2}$ into a slower exponential gives the last inequality.
\\

\noindent
Interior-volume and Gaussian-tail bounds give $W_n^\Gamma(s)\leq C$ for $s\geq\ell^2$, ensuring large-time integrability at $m>0$.
The remainder and $\OO(s)$ mass correction are $ds/s$-integrable near $s=0$, proving Eq.~\eqref{eq:additive}.
Global boundedness, the small-time limit, and the majorant $C|\delta\chi(u)|/u$ also justify the cutoff argument in Appendix~\ref{sec:app-cutoff} for a fixed graph and operator.

\section{Topology of Link}
\label{sec:appF}
\noindent
Each completed branch point is locally $z\mapsto z^n$, so the connected link is a closed orientable surface of genus $g_n$ from Eq.~\eqref{eq:genus}.
For its completed infinite cone $X$ with apex $p$, radial contraction, projection, and the pair's exact homology sequence give
\begin{equation}
H_2(X,X\setminus\{p\};\mathbb Z)
\cong H_1(\Sigma_n^{(3)};\mathbb Z)=\mathbb Z^{2g_n}.
\label{eq:app-local-homology}
\end{equation}
Here $H_k$ denotes degree-$k$ integer homology.
At a three-manifold interior point, the relative group vanishes; hence the apex is nonmanifold for $n\geq3$.
For $n=2$, the cone is topologically $\mathbb R^3$ with a singular metric.
\section{Convergent Spectral Formula on Replica Link}
\label{sec:appH}
\noindent
We choose a uniform geodesic radius whose balls contain at most one branch value; selected bridges must exit their base-point balls.
First-exit conditioning before $t/2$, time reversal, and Gaussian heat-kernel bounds give $0\leq F_n(t)\leq Ct^{-1}\exp(-c/t)$ for $0<t\leq1$.
The constants depend on the positive angles, not $n$.
Together with the zero-mode limit given in the Letter, this proves endpoint integrability in Eq.~\eqref{eq:spectral}.
\\

\noindent
For $\nu=\sqrt{\lambda+1/4}\geq1/2$, setting $f=r^{-1/2}u$ preserves $\int r^2|f|^2dr=\int r|u|^2dr$ and gives radial operator $-\partial_r^2-r^{-1}\partial_r+\nu^2/r^2$.
We define
\begin{equation}
I_\nu(z)=\sum_{q=0}^{\infty}\frac{(z/2)^{2q+\nu}}{q!\,\Gamma(q+\nu+1)},\qquad z>0.
\label{eq:app-bessel-definition}
\end{equation}
Here $\Gamma(a)=\int_0^\infty y^{a-1}\exp(-y)dy$, $a>0$; $J_\nu$ is the same series with each term multiplied by $(-1)^q$.
The Friedrichs modes are $J_\nu(kr)$ with momentum $k>0$; the normalization $\int_0^\infty k\,dk\,J_\nu(kr)J_\nu(kr')=\delta(r-r')/r$, with Dirac delta $\delta$, gives, relative to $r'^2dr'$,
\begin{align}
k_\nu(r,r';s)
&=(rr')^{-1/2}\int_0^\infty k\,dk\,e^{-sk^2}J_\nu(kr)J_\nu(kr')\nonumber\\
&=\frac{e^{-(r^2+r'^2)/(4s)}}{2s\sqrt{rr'}}I_\nu\!\left(\frac{rr'}{2s}\right),\quad r,r',s>0.
\label{eq:app-cone-radial-kernel}
\end{align}
Termwise Gaussian integration proves the second equality; $I_\nu(z)\leq(z/2)^\nu \exp(z)/\Gamma(\nu+1)$ ensures absolute convergence.
\\

\noindent
For $\mathscr L_\nu(\beta)=\int_0^\infty \exp(-\beta z)I_\nu(z)dz$, $\beta>1$, Bessel's equation gives $(\beta^2-1)\mathscr L_\nu''+3\beta\mathscr L_\nu'+(1-\nu^2)\mathscr L_\nu=0$, with primes denoting $\beta$ derivatives.
With $\cosh\alpha=(\exp(\alpha)+\exp(-\alpha))/2$ and $\sinh\alpha=(\exp(\alpha)-\exp(-\alpha))/2$, setting $Y(\alpha)=\sinh\alpha\,\mathscr L_\nu(\cosh\alpha)$ gives $d^2Y/d\alpha^2=\nu^2Y$; $\mathscr L_\nu(\beta)\sim2^{-\nu}\beta^{-\nu-1}$ fixes $Y=\exp(-\nu\alpha)$ and proves Eq.~\eqref{eq:radial-trace}.
The covering bound $K_\Sigma(t)\leq d_\Sigma K_{S^2}(t)=\OO(t^{-1})$ as $t\downarrow0$, with covering degree $d_\Sigma$, ensures finite damped spectral sums.
Subordination follows from the Gaussian integral
\begin{equation}
\int_0^\infty t^{-3/2}e^{-\alpha^2/(4t)-\zeta t}dt
=\frac{2\sqrt\pi}{\alpha}e^{-\alpha\sqrt\zeta},\qquad\alpha,\zeta>0.
\label{eq:app-subordination}
\end{equation}
We apply this identity to each positive spectral sum before taking their finite combination.
The endpoint bounds justify the limit in Eq.~\eqref{eq:spectral}.

\end{document}